# A review of heat transport in solvated gold nanoparticles: Molecular dynamics modeling and experimental perspectives

Md Adnan Mahathir Munshi[1], Emdadul Haque Chowdhury[1], Luis E. Paniagua-Guerra[1], Jaymes Dionne[2], Ashutosh Giri[2], Bladimir Ramos-Alvarado[1, *]

[1]Department of Mechanical Engineering, The Pennsylvania State University, University Park, Pennsylvania-16802, USA.

[2]Department of Mechanical Engineering, University of Rhode Island, Kingston, RI 02881, USA.

## Abstract

Turning gold nanoparticles (AuNPs) into nanoscale heat sources via light irradiation has prompted significant research interest, particularly for biomedical applications over the past few decades. The AuNP's tunable photothermal effect, notable biocompatibility, and ability to serve as vehicles for temperature-sensitive chemical linkers enable thermo-therapeutics, such as localized drug/gene delivery and thermal ablation of cancerous tissue. Thermal transport in aqueous AuNP solutions stands as the fundamental challenge to developing targeted thermal therapies; thus, this review article surveys recent advancements in our understanding of heat transfer and surface chemistry in AuNPs, with a particular focus on thermal boundary conductance across gold- and functionalized-gold-water interfaces. This review article highlights computational advances based on molecular dynamics simulations that offer valuable insights into nanoscopic interfacial heat transfer in solvated interfaces, particularly for chemically functionalized AuNPs. Additionally, it outlines current experimental techniques for measuring interfacial thermal transport, their limitations, and potential pathways to improve sensitivity. This review further examines computational methodologies to guide the accurate modeling of solvated gold interfaces. Finally, it concludes with a discussion of future research directions aimed at deepening our understanding of interfacial heat transfer in solvated AuNPs, crucial to optimize thermoplasmonic applications.

*Corresponding Author. E-mail address: bzr52@psu.edu

## 1. Introduction

The utilization of gold nanoparticles (AuNPs) in the biomedical field has gained significant momentum in recent years due to their unique characteristics, such as tunable optical properties, and remarkable biocompatibility.[1,2] Upon light irradiation, free electrons in noble metal nanoparticles undergo oscillations,[3,4] creating localized surface plasmons resonance (LSPRs), which enables efficient light absorption, scattering, and focusing.[5,6] Metallic NPs with LSPR, known as plasmonic NPs, act as potent nanoscale heat sources. By tuning the absorption rate relative to the cooling rate, their thermal response can be optimized to meet specific requirements, giving rise to thermoplasmonics.[7] As a result, their applications include photothermal reaction acceleration,[8] chemical catalysis,[9] solar energy harvesting,[10] additive manufacturing,[11] and thermal sensing at solid-liquid interfaces.[12] By enabling precise interactions with biological systems at the molecular level, thermoplasmonics enable treatments such as photothermal ablation therapy for cancer,[13] bacterial eradication,[14,15] and advanced drug delivery systems capable of co-delivering anticancer agents,[16,17] such as RNA, DNA,[18–20] and proteins.[21,22] Consequently, a successful thermoplasmonic implementation requires a deep understanding of the thermal response of heated plasmonic NPs.

Effective plasmonic NP-based therapy and drug delivery require optimal size and morphology for cellular uptake, high biocompatibility, and efficient heat generation while minimizing damage to healthy tissues. Effective bioincorporation requires precise NP delivery to target tissues, ensuring selective accumulation. In passive targeting, the enhanced permeability and retention effect facilitates NP accumulation in tumor tissues with leaky vasculature.[23–25] When passive targeting is insufficient, active targeting is employed by functionalizing NPs with specific ligands for selective receptor binding.[26–29] Post-delivery, adverse effects are mitigated by utilizing laser wavelengths within biological windows (700–980 nm and 1000–1400 nm),[30,31] where tissue absorption and scattering are minimized due to enhanced optical transparency. Consequently, plasmonic NPs in biomedical applications primarily operate within the visible and near-infrared spectrum.[32]

Among the materials explored for photothermal effect-based biomedical applications, gold, silver, and copper stand out due to their LSPR-spanning wavelengths, which allow for adequate tissue penetration.[33] While silver has a higher photothermal conversion efficiency than gold, both silver and copper exhibit significant toxicity and lack the chemical stability required for in vivo applications.[34,35] As a result, AuNPs are preferred due to their chemical stability, low cytotoxicity, and versatile functionalization capabilities,[36] i.e., their ability to undergo surface functionalization via strong sulfur-gold bonds, enhancing biocompatibility.[37,38] Thiol functionalization enables AuNPs to conjugate therapeutic molecules, targeting ligands, and passivating agents, thereby improving in vivo stability and biological interactions.[13,39]

Thiol-functionalized AuNPs have been proposed as drug delivery vehicles, enabling targeted drug release via bond cleavage under reducing conditions.[26,40–42] They have been extensively studied for nucleic acid delivery via click chemistry,[42–46] a method that provides precise temporal drug release while maintaining biocompatibility and minimizing biological disruption.[47,48] Diels-Alder (DA) reactions (click chemistry) are widely used to form stable cyclohexene derivatives through a reaction between a conjugated diene and a dienophile.[49,50] At elevated temperatures, the reaction can reverse via the retro-Diels-Alder (rDA) pathway, regenerating the original diene and dienophile products (see Fig. 1).[50,51] The thermal response of the diene/dienophile linkers can be fine-tuned to achieve controlled drug release, enabling temporal delivery of multiple drugs. Thus, understanding the thermal behavior of plasmonic NPs becomes crucial for optimizing drug delivery systems.

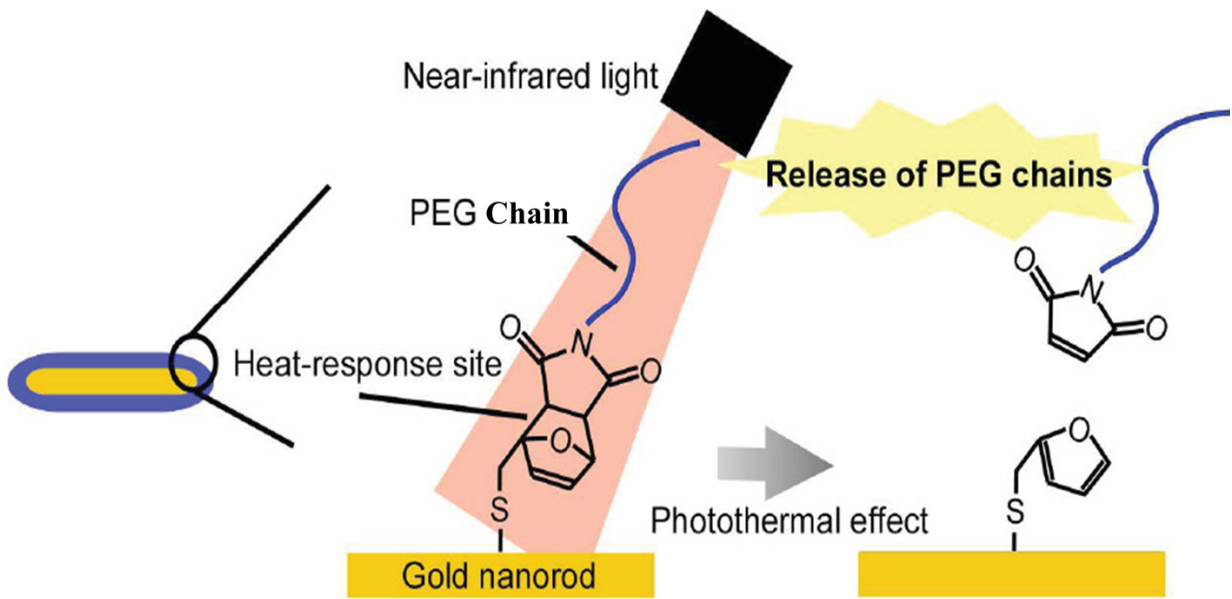

**Fig. 1** Near-infrared laser light irradiation-induced photothermal heating triggers retro Diels-Alder cleavage of the surface of the PEG-DA-modified gold nanorods, releasing PEG and causing gold nanorod aggregation. (this figure has been adapted from ref. 51 with permission from ACS publications, Copyright 2011).

Advancing thermoplasmonics requires a deep understanding of heat transport at the NP-solvent interface, which is quantified by the thermal boundary conductance (TBC). Thermal transport between NPs and their surrounding environment is influenced by several factors such as composition,[52] size,[53–56] surface properties,[12,57,58] the dynamics of the solvent molecules,[59] solid-liquid affinity,[12] their mobility,[60] and the density of covalent bonds at the NP surface[61] (see Section 2 for an extended discussion). Ligands functionalization of plasmonic NPs adds complexity due to the formation of a three-component interface comprising metal, ligands, and solvent. Independently of the interface's composition, the TBC is essential for temperature control in thermoplasmonics. Accurate local temperature measurements in nanomaterials remain difficult, underscoring the need for advanced tools to characterize heat dissipation in plasmonic NP systems. Earlier investigations used continuum heat transfer models to correlate time-dependent NP temperature changes with heat flow, these models have shown limited success.[62] In contrast, atomistic simulations have become a powerful alternative, offering high-resolution insights and greater flexibility for investigating nanoscale thermal transport.[63,64]

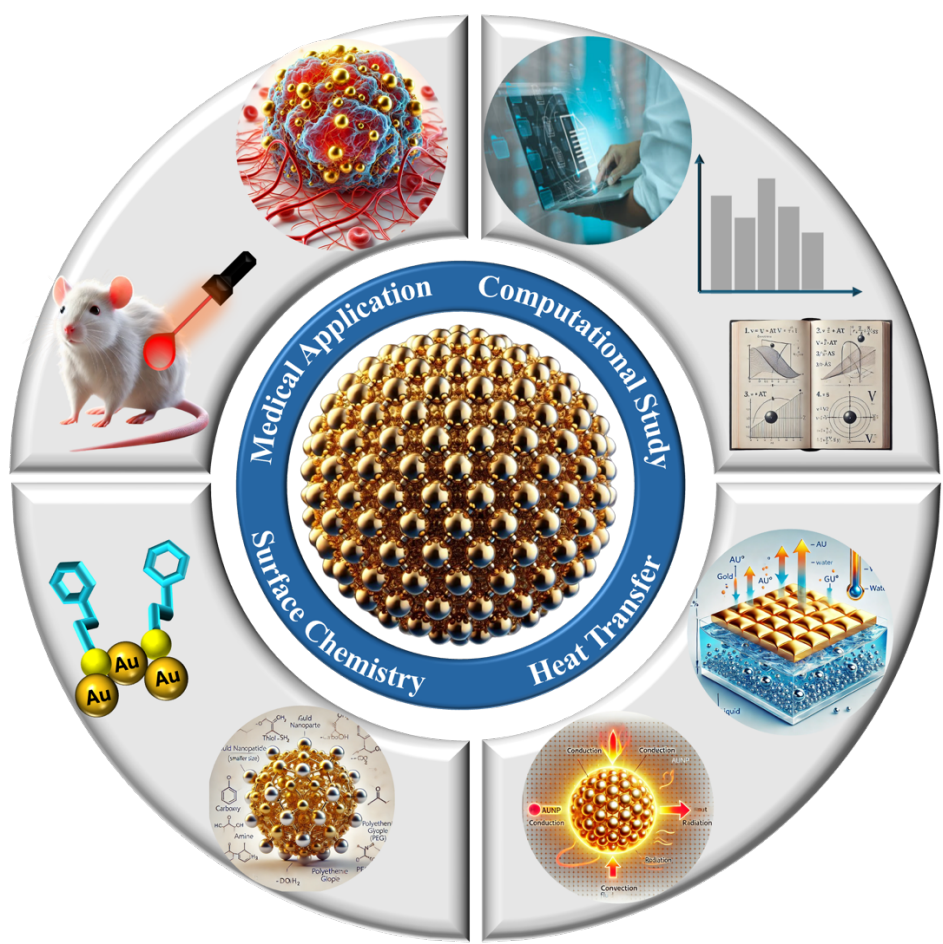

**Fig. 2** Knowledge map of the literature review on solvated AuNPs.

As previously indicated, the spatiotemporal temperature control of plasmonic NPs depends on fine-tuning the photothermal effect and the particle-solvent heat dissipation. This review article surveys the recent literature on the latter. Section 2 reviews the literature on the fundamental physics and mechanisms governing heat transfer across solid-liquid interfaces, with a focus on AuNPs systems and interfaces functionalized with a variety of ligands and self-assembled monolayers (SAMs). Section 3 discusses recent advances in experimental techniques for characterizing thermal transport across solid-liquid interfaces, their limitations, and emerging alternatives to improve the sensitivity of measurements. Section 4 outlines the computational methodologies and models available for simulating functionalized gold-water interfaces, highlighting their role in predicting interfacial thermal transport. Finally, the review concludes in Section 5 with a summary of key findings and an outlook on future research directions. Fig. 2 presents a knowledge map outlining the structure and scope of this review.

## 2. Thermal Transport in Solvated Metal Nanoparticles

The study of thermal transport across solid-liquid interfaces and the concept of TBC traces back to Pyotr Kapitza's investigation of the thermal conductivity of helium capillaries.[65] Kapitza observed a temperature discontinuity between heated metallic surfaces and liquid helium, leading to the idea that this discontinuity is proportional to the heat flux across any heated surface. Heat transfer across the interface between different materials inherently encounters resistance due to the abrupt change in thermal properties; this is known as thermal boundary resistance (TBR). The inverse of TBR is the TBC, defined as $J=G\Delta T_{int}$, where $J$ represents the heat flux across the interface, $\Delta T_{int}$ is the temperature discontinuity at the interface, and $G$ denotes the TBC. During the latter half of the 20th century, research focus shifted from solid-liquid to solid-solid interfaces, driven by advancements in microelectronics.[66] In Cahill et al.'s seminal review on nanoscale heat transfer, solid-liquid heat transfer was not explicitly addressed;[67] however, a follow-up review indicated a growing interest in solid-liquid interfaces between 2002 and 2012.[68] Similarly, Luo and Chen[69] recognized their increasing significance in biomedical applications, catalysis, energy generation, and colloidal suspensions. The early 21st century saw a renewed focus on solid-liquid TBC, particularly following the work of Ge et al.,[70] who measured thermal transport at hydrophilic and hydrophobic interfaces, underscoring the need for a deeper understanding of interfacial heat transfer mechanisms.

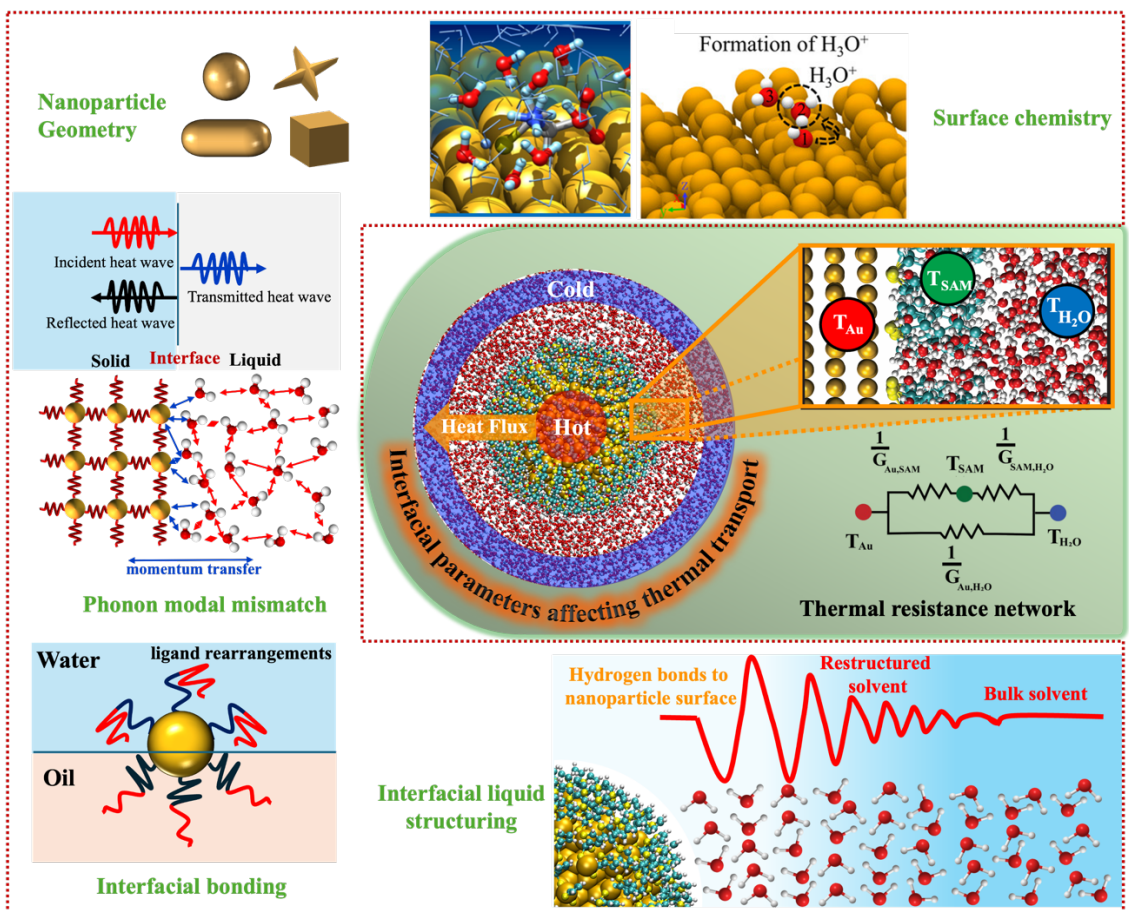

**Fig. 3** Interfacial parameters affecting thermal across solid-liquid interfaces and its adaptation to AuNP-water interfaces. Interfacial interaction between AuNP and water (top-right); Nanoparticle size can affect interfacial chemistry (top-left); Phonon modal mismatch (left-middle); Interfacial bonding between AuNP and ligand in oil and water (bottom-left); Enhanced short-range order of solvent molecules at NP surfaces (bottom-right).

Extensive research indicates that the TBC across solid-liquid interfaces is governed by a complex interplay of factors: (i) the nature of the bonds at the interface, (ii) the interfacial liquid structure, (iii) the strength of atomic interactions across the interface, and (iv) vibrational mismatch between the solid and liquid phases, see Fig. 3 for a graphical depiction of these factors. Despite these advancements, the mechanisms governing solid-liquid thermal transport remain only partially understood, necessitating interdisciplinary efforts to uncover the underlying principles. This complexity is further heightened in solvated NPs, where the morphology of the NP must be considered, and clustering can occur. Moreover, when organic ligands or polymers are used to functionalize solid surfaces, the dynamics of the atoms at the interface and thus the heat transfer mechanisms are modified. Accordingly, this section is organized into three subsections: Section 2.1 reviews the fundamentals of interfacial thermal transport at solid-liquid interfaces. Section 2.2 focuses on specific effects observed in solvated NP systems, and Section 2.3 examines the additional complexities associated with characterizing thermal transport across functionalized interfaces.

**2.1 Fundamentals of Solid-Liquid Interfacial Thermal Transport**

A consensus from early research is that the solid-liquid affinity, often quantified by the equilibrium contact angle ($\theta_c$), is the key parameter controlling the interfacial heat transfer. Ge et al.[70] reported early measurements of the TBC across solid-water interfaces and observed lower TBC values in the range of 50-60 MW/m$^2$K for hydrophobic surfaces, compared to values of 100-180 MW/m$^2$K for hydrophilic surfaces. These observations led to the argument that hydrophilic surfaces attract a higher density of liquid molecules near the interface, thereby increasing the energy carriers available for heat transfer near the interface. Consecutively, TBC-wettability relationships were sought after and developed. Early molecular dynamics (MD) investigations[71–73] showed that the TBC increases sharply as the solid-liquid bonding strength increases, approaching a finite value in the limit of complete wetting, corresponding to a contact angle of $\theta_c$ = 0°. Shenogina et al.[74] further identified a scaling law of the form $G \sim 1+ \cos(\theta_c)$, relating it to the work of adhesion ($W_{ad}$), where $W_{ad} = \gamma_{lv}[1 + \cos(\theta_c)]$, with $\gamma_{lv}$ being the liquid-vapor surface tension (see Fig. 4(a)). Subsequent investigations[75–77] confirmed the $G \sim 1 + \cos(\theta_c)$ scaling law, and Alexeev et al.[75] further suggested that it could be general across different interfaces.

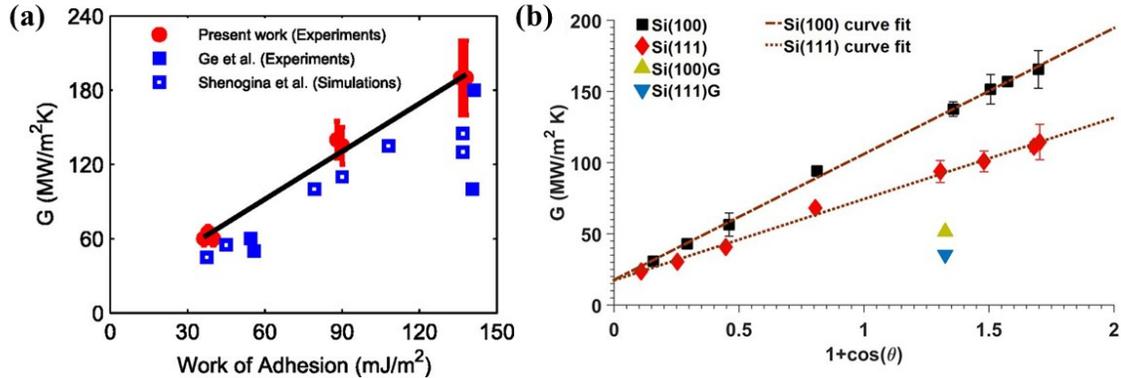

**Fig. 4** Descriptions of the TBC using wettability. (a) Early contributions suggested a quasi-universal law of the form $G \sim 1 + \cos(\theta_c)$, from a relationship to the work of adhesion $W_{ad} = \gamma_{lv}[1 + \cos(\theta_c)]$ (Reprinted with permission from ref. 77. Copyright 2013 AIP publishing). Challenges to the TBC-wettability notion. (b) TBC across silicon- and graphene-coated-silicon-water surfaces. (Reprinted with permission from ref. 79. Copyright 2016 ACS publication.)

The applicability of quasi-universal TBC-wettability relationships has been challenged due to the oversight of the complex interplay of interfacial mechanisms dictating solid-liquid heat transfer. For example, Acharya et al.[78] highlighted the limitations of using the contact angle on superhydrophilic surfaces. Similarly, at interfaces with large curvatures, where an accurate determination of the contact angle is difficult, TBC-wettability relationships become limited (see Section 2.2 for an extended discussion on curvature effects). Furthermore, the generality of the scaling relationship $G \sim 1 + \cos(\theta_c)$ has been challenged by Ramos-Alvarado et al.[79–81] revealing two paradigm shifts: first, more wettable crystallographic planes can be less conductive; second, in such cases, $G \sim 1 + \cos(\theta_c)$ holds independently for each plane but lacks universality (see Fig. 4(b)). Subsequent work demonstrated that factors such as chemical composition,[82] crystallographic structure of the solid surface,[83,84] and interface curvature[85] can undermine the quasi-universal nature of the TBC-wettability relationship. These contradictions to the early TBC-wettability relationship highlight the intricate nature of heat transfer at solid-liquid interfaces, emphasize the need to look beyond the solid-liquid interaction strength and underscore the importance of exploring additional factors influencing TBC behavior.

The nature of the interfacial bonds plays a crucial role in determining the TBC. For instance, several investigations have reported an increased TBC across various solid-liquid interfaces due to polarization effects,[86–90] which are crucial for metallic NPs immersed in biological environments.[91] In these cases, polarization causes negligible changes in wettability and interfacial free energy, but it influences the molecular ordering of the liquid phase, which has been attributed to enhancing the TBC of polar interfaces due to favoring the formation of hydrogen bonds (H-bonds).[82,88,89,92] The mainstream perspective suggests ordered H-bonds pull liquid molecules closer to the solid surface, enhancing the TBC. Alternatively, the TBC enhancement in polarizable interfaces has been explained by the excitation of additional degrees of freedom, such as vibrational modes in polar solvents.[86] Nevertheless, it has been reported that polarizability negligibly modifies the vibrational density of states (vDOS) of the metal or liquid phases.[89] In such cases, the enhancement in TBC has been attributed to a rise in the phonon transmission probability at the interface, which is sensitive to molecular ordering and interatomic spacing.

Vibrational mode mismatch between solid and liquid particles is a key factor governing phonon-mediated thermal transport across interfaces, where the overlap of the phonon density of states (DOS) between the two phases quantifies this effect.[93–97] Giri and Hopkins[98] used simple Lennard-Jones (LJ) solid-liquid MD models to show that stronger solid-liquid bonding enhances low-frequency phonon coupling, broadens the interfacial DOS, and introduces new phonon modes;

thus, increasing the TBC. In contrast, weak or hydrophobic interfaces act like free surfaces, limiting phonon transmission. Han et al.[99] reported a similar shift of vibrational modes to higher frequencies in perfluorohexane, though driven by increased liquid pressure rather than bonding. Surface functionalization like self-assembled monolayers (SAMs) and chemical passivation can reduce the vibrational mismatch by introducing buffer interfacial modes, improving phonon overlap even when the bulk DOS differs.[100–103] However, the relationship between modal overlap and TBC is not universal; interfacial liquid structuring and the directionality of heat flux (in-plane vs. out-of-plane) also influence transport. Out-of-plane modes dominate at low-affinity interfaces, while strong bonding and ordered structuring (e.g., via hydrogen bonding or electrostatics) enhance in-plane contributions.[104–108] These findings highlight the need to consider vibrational mismatch, interfacial chemistry, and liquid ordering for a comprehensive understanding of nanoscale heat transfer across solid-liquid interfaces.

The role of the liquid's molecular interfacial organization in the TBC has been extensively investigated. Several authors have shown that liquid layering at the interface is essential in determining the TBC. The prevailing hypothesis is that since molecules are the primary energy carriers in liquids, their availability and proximity to the interface significantly impact the energy transfer probability. Furthermore, microcalorimetry and heat capacity measurements have revealed that absorbed water on metal oxide surfaces exhibits distinct thermodynamic properties compared to bulk water,[109] suggesting enhanced thermal properties for interfacial liquids. Early MD investigations linked the TBC to the height and location of the first hydration layer near the interface,[75,87,88,110] showing that higher TBC values correlate with enhanced liquid layering, although this relationship is non-universal.[88] Subsequent works investigated the complex liquid layering that extends beyond the first hydration layer,[111,112] while also accounting for interfacial pressure effects[112] and the formation of solid-like structures in the liquid phase.[113] Recent work by Paniagua et al.[85] further underscores the importance of interfacial liquid organization, showing that the TBC is significantly enhanced when liquid molecules near the interface organize into cluster-like structures, in contrast to uniform layers.

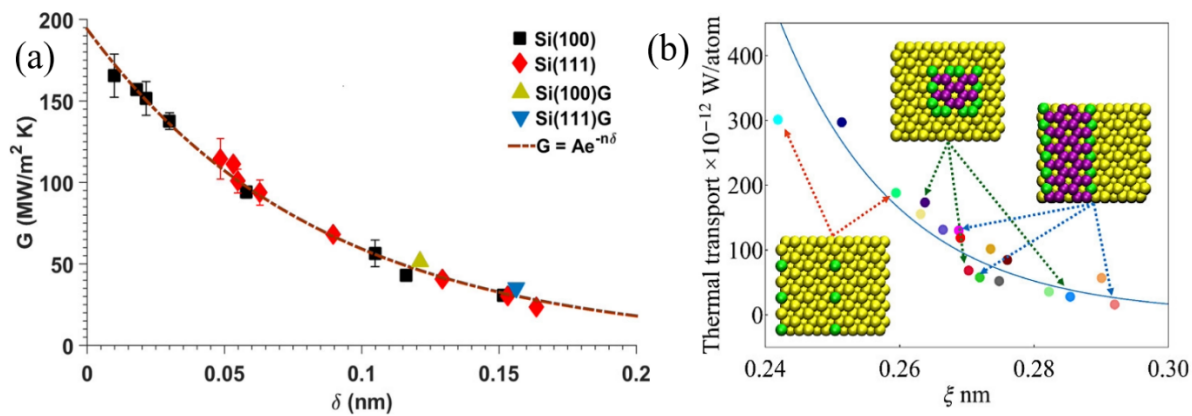

**Fig. 5** (a) Reconciliation of the anisotropic TBC calculated for different surfaces in contact with water using the density depletion length $\delta$. (this figure has been adapted from ref. 79 with permission from ACS Publications, Copyright 2016). (b) TBC as a function of the radial density depletion length $\xi$. (this figure has been adapted from ref. 114 with permission from ACS Publications, Copyright 2024).

Building on the observed dependence of TBC on interfacial liquid layering, Ramos-Alvarado et al.[79] used the density depletion length $\delta$ as a parameter to reconcile the anisotropic TBC calculations of silicon-water interfaces as illustrated in Fig. 5(a). More recently, Motokawa et al.[114] introduced the radial density depletion length (RDDL) to account for single-atomic structures on solid surfaces, demonstrating that liquid ordering significantly modulates interfacial thermal transport as illustrated in Fig. 5(b). The concept of $\delta$, which quantifies the deficit or surplus liquid molecules near the interface, has also been employed to describe hydrodynamic slip.[115–118] Subsequent contributions validated $\delta$ as a reliable parameter for describing the TBC across different interfaces. [82,84,85,108]

## 2.2 Thermal Transport in Solvated Nanoparticles

Thermal transport measurements across individual NPs are challenging; thus, several experimental contributions have focused on macroscale insights into the thermal relaxation of colloidal NP solutions also known as nanofluids. These works have revealed dependencies on the intensity of optical excitation[119] and NP concentration.[120,121] Additionally, enhancements in the thermal conductivity of nanofluids have been shown to depend on NP size.[122] To understand the mechanisms driving these enhancements, MD models have focused on computing the effective thermal conductivity (ETC) of nanofluids. Sarkar and Selvam[123] demonstrated that while the diffusion of NPs is slower than that of liquid atoms, the interfacial liquid atoms surrounding the NPs exhibit enhanced movement compared to bulk liquid atoms, which contributes to increased ETC. Further research has examined the impact of NP aggregation, reporting that aggregation enhances the ETC of nanofluids, with chain-like NP aggregates providing a greater increase in ETC compared to spherical aggregates.[124,125] Liquid layering at NP interfaces has also been explored, suggesting that the local ETC of liquid adsorption layers increases with NP wettability,[126] shedding light on the role of interfacial phenomena in thermal transport within nanofluids.

A deeper understanding of thermal transport across curved interfaces can be achieved by focusing on the TBC at the NP interface rather than the ETC. Fundamentally, the cooling dynamics of rapidly heated NPs can be estimated by assessing the TBC and the NP's size. In their experimental work, Ge et al.[59] showed that for sufficiently large spherical NPs or high TBC interfaces, the NP's temperature decay is limited by thermal diffusion in the surrounding fluid. The characteristic diffusion time ($\tau_d$) can be estimated by equating the particle's heat capacity with that of the fluid within a thermal diffusion length. Conversely, for sufficiently small NPs or low

TBC interfaces, the cooling rate is limited by the TBC, with the characteristic decay time ($\tau_i$) determined by the ratio of the particle's heat capacity to the total interfacial thermal conductance. Based on this, Ge et al.[59] proposed a critical TBC value,

$$\text{TBC}_c = \frac{3 C_f \Lambda_f}{r_p C_p} \quad (1)$$

where $\Lambda_f$ is the thermal conductivity of the fluid, $C_f$ and $C_p$ are the volumetric heat capacities of the fluid and NP, respectively, and $r_p$ is the NP radius. This formulation effectively demarcates two distinct cooling regimes: when TBC $\gg$ TBC$_c$, the cooling is diffusion-limited ($\Lambda_f C_f$); when TBC $\ll$ TBC$_c$, it is interface-limited. More recently, Wilson et al.[127] proposed an alternative definition of the critical TBC, grounded in the concept of the Kapitza length (effective length creating the same thermal resistance as an interface). They defined the critical conductance for water-solvated particles as,

$$G_c = \frac{2 k_w}{d} \quad (2)$$

where $k_w$ is water's thermal conductivity, and $d$ is the NP's diameter. Similar to Ge et al.'s[59] conclusions: If $G \gg G_c$ the NP's cooling process is dominated by water diffusion; alternatively, if $G \ll G_c$ the NP's cooling is controlled by the interface. Wilson et al.[127] defined the diffusion-dominated regime as $G > 10 G_c$, the interface-dominated regimes as $G < 0.1 G_c$, and a mixed regime as $0.1 G_c < G < 10 G_c$. These regions, along with a survey of experimental and computational data are plotted in Fig 6(a), where it can be observed that most AuNP systems exist in the mixed regime and skew towards the interface-dominated area; thus, supporting the need for further research on interfacial heat transfer in solvated AuNPs.

A 10 nm AuNP in water can be considered as an example to contrast the two critical TBC models. In this situation, critical TBCs of 300 MW/m²K and 100 MW/m²K are obtained using Eq.

(1) and Eq. (2), respectively, and while these numbers differ by a factor of three, they exist in the same cooling regime per Fig. 6(a). The difference lies in the fact that Eq. (1) was derived by obtaining the ratio of the thermal time constants in the particle and surrounding liquid, i.e., transient heat transfer parameters, and Eq. (2) was derived using a Kapitza conduction length analogy. Notably, Ge et al.'s[59] model is more conservative if one were to follow the same mapping strategy as Wilson et al.[127] depicted in Fig. 6(a), strengthening the call for a deeper fundamental understanding of interfacial heat transfer in solvated NP systems.

NP size and curvature are synonyms of the same parameter that can be computationally investigated. Merabia et al.[128,129] used MD simulations of solvated AuNPs to demonstrate that curvature significantly alters the thermodynamic properties of the interfacial liquid. Due to their curved geometry, spherical AuNPs could be heated above their melting temperature without causing a phase change in the adjacent liquid. Additionally, a vapor layer, which typically forms on flat interfaces under similar heating conditions, was notably absent at curved spherical interfaces. The delay in liquid phase change and AuNP melting was attributed to the extremely high pressure near the curved interface, i.e., the Laplace pressure generated by the NP's curvature. Later, in their computational work on nanoscale boiling around AuNPs, Gutiérrez-Varela et al.[130] demonstrated that the formation of a low-density liquid layering during heating transiently reduces the TBC and delays vapor nanobubble onset at the AuNP-water interface, which also explains the absence of interfacial water phase change by Merabia et al.[128] Notably, when evaporation was reached, it was reported that nanobubbles nucleate more rapidly on hydrophilic nanoparticles, contrary to predictions from isothermal classical nucleation theory. Merabia et al. and Gutiérrez-Varela et al.[128–130] contributions demonstrate the potentially devastating effects of poor spatiotemporal temperature control of AuNP therapies. While particle melting and waster nucleation could be delayed due to the large Laplace pressure around spherical NPs, evaporation is still plausible, and its subsequent effects are lower TBCs and eventually potential AuNP melting. Thus, particle size effects and interfacial liquid structure properties must be better understood to engineer AuNP's temperature controls.

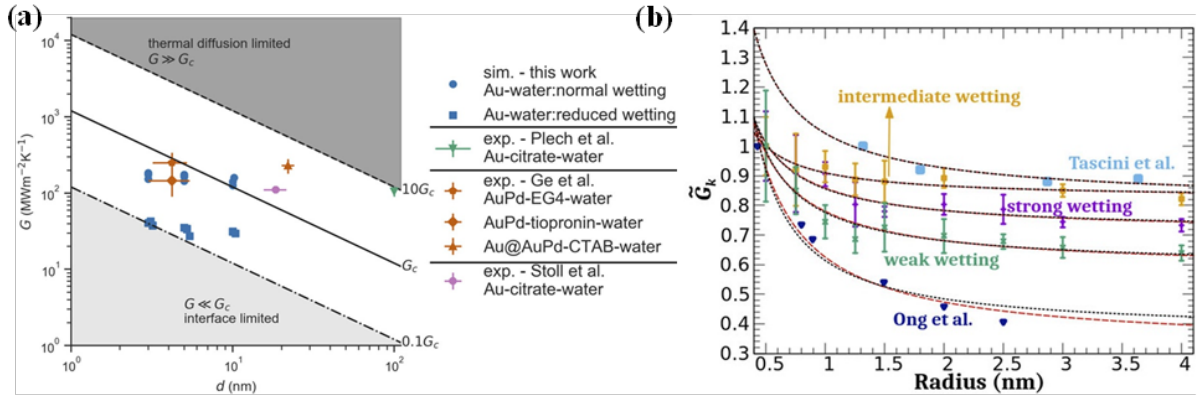

**Fig. 6** (a) Impact of the TBC on the nanoparticle's spatiotemporal temperature regulation (this figure has been adapted from ref. 127 with permission from AIP Publishing, Copyright 2022). (b) Normalized interfacial thermal conductance of nanoparticles as a function of the nanoparticle radius. (this figure has been adapted from ref. 131 with permission from AIP Publications, Copyright 2022).

The effect of curvature on the TBC across NP-liquid interfaces was further explored by Tascini et al.,[56] using a generic MD model. They demonstrated a strong dependence of the TBC on the NP's curvature, showing that the TBC increases with interfacial curvature across a wide range of fluid-solid interaction strength. Their findings revealed an empirical relationship described by $G = G_\infty + c/r$, where $1/r$ is the NP's curvature, $G_\infty$ is the TBC in the limit of $r$ going to infinity or a flat surface, and $c$ is a fitting parameter. They observed that stronger interfacial interactions lead to larger values of $c$. Building on this, Gutierrez-Varela et al.[131] investigated the impact of curvature and size on the TBC of AuNPs across three distinct wetting regimes-strong, intermediate, and weak-illustrated in Fig. 6(b). Their calculations matched the curvature effect of Tascini et al.'s[56] empirical correlation but using a realistic metal-liquid system. Gutierrez-Varela et al.[131] explained the enhanced conductance of smaller AuNPs using two arguments. First, smaller AuNPs have higher solid-liquid coordination numbers (a greater number of water molecules per surface atom), which creates a higher water-Au potential energy. Second, the NP's curvature alters the interfacial vibrational spectrum: as the AuNP size decreases, the high-frequency van Hove peak fades while the low-frequency peak strengthens, aligning Au and water vibrations more closely and enhancing the TBC. Additionally, they observed that for smaller NPs, the amplitude of the first peak in the water density profile increases, consequently enhancing the structuring of interfacial water. Yet, they caution that the correspondence between interfacial conductance and fluid density is not universal.

Expanding on these insights, Paniagua-Guerra and Ramos-Alvarado[85] investigated interfacial heat transfer at AuNP-water interfaces, emphasizing the role of the density depletion length ($\delta$). Their MD simulations demonstrated that curved interfaces consistently exhibited higher TBC than flat surfaces. This enhancement was attributed to the larger availability of water molecules at the interface, which facilitated energy transfer. Additionally, they identified an exponential relationship between the TBC and $\delta$, indicative of the transferability of the TBC-$\delta$ relationship to curved interfaces, where traditional wettability metrics are difficult to compute.

The influence of NP morphology on TBC has been further explored by studying NPs with various shapes. Neidhart and Gezelter[132] dispersed bare AuNPs- icosahedral, cuboctahedral, and spherical- in solvent and systematically examined how NP morphology influences the TBC by quantifying the density of undercoordinated sites on the solid surface. They observed higher TBC values for particles with a greater fraction of exposed undercoordinated atoms. Building on this concept, Jiang et al.[133] showed that TBC can vary locally across an NP's surface: solid atoms with lower coordination numbers- i.e., fewer neighboring atoms-make more contact with the solvent, enhancing local heat transfer. Similarly, Gutierrez-Varela[131] quantified the number of water molecules interacting with a surface gold atom, via the water-gold potential energy, and demonstrated that decreasing the NP's size, increases this number, thereby enhancing TBC. These findings emphasize the critical role of NP shape and atomic coordination in determining interfacial thermal transport properties.

In summary, the strong dependence of thermal boundary conductance (TBC) on morphological factors highlights the complexity of describing interfacial thermal transport at solid-liquid interfaces. This complexity extends beyond a simple characterization of interfacial bonding strength, as the energy landscape at the interface is influenced by both the solid surface morphology and the strength of interfacial energy interactions. The analysis becomes increasingly intricate as the interface structure grows more complex. However, the underlying mechanisms and physics governing interfacial thermal transport remain consistent, Consequently, much of the knowledge gained from thermal transport across bare solid-liquid interfaces can be applied to functionalized solid-liquid interfaces, as will be explored in detail in the following Subsection.

## 2.3 Thermal Transport at Functionalized Solid-Liquid Interfaces

Interfacial heat transport in functionalized AuNPs is strongly influenced by the ligand type and surface morphology and can be easily represented by the resistance network panel in Fig 3. The dependence of interfacial thermal transport on ligand-solvent interactions was initially highlighted by Ge et al.[59,70] who demonstrated that the affinity between the solvent and the terminal groups of ligands significantly influences the TBC. More recent investigations have further demonstrated that the pronounced enhancement in TBC arises from tight coupling at two interfaces: from the metal core to the ligand shell, and from the ligands to the surrounding fluid. For instance, Au surfaces functionalized with polyethylene glycol (PEG) exhibit significantly higher TBCs in water compared to those functionalized by citrate or cetyltrimethylammonium bromide (CTAB) ligands.[134] This enhancement arises from the strong Au-S bonds that couple the Au core to the PEG ligands, adding to the increased physical contact between those ligands and the surrounding solvent. Similarly, the presence of a ligand layer that structurally/chemically matches the solvent can create a buffer layer that reduces the vibrational modal mismatch at the interface.[135] Lastly, the ligand surface coverage could increase the availability of channels for interfacial conduction.[136–138] These three individual effects on the TBC will be discussed further in this Section.

For heat to flow from a functionalized solid to a solvent, efficient transport occurs from the solid to the ligands due to strong covalent chemical bonds. This is followed by heat transfer from the ligands to the liquid solvent, which could be enabled via vibrational coupling.[135,138–141] The ligand layer may act as an intermediary, bridging the solid and liquid phases, which typically exhibit significant vibrational mismatches.[136,138,140] For instance, Kikugawa et al.[138] demonstrated via vibrational analysis that self-assembled monolayers on Au significantly reduce interfacial thermal resistance compared to bare gold-solvent interfaces. Similarly, Hannah and Gezelter[136] showed that hexylamine ligands enhanced vibrational overlap between CdSe and hexane (the surrounding solvent), thereby facilitating improved interfacial heat dissipation. Contrariwise, Hung et al.[101] reported a negligible phonon spectral overlap effect in SAM-coated gold and water, where better vibrational coupling did not correlate with higher TBC. Instead, they found that thermal transport is primarily facilitated by the aggregation of water molecules around the terminal atoms of SAM. So, depending on the vibrational properties of both the liquid solvent and the ligand layer, the ligand-liquid interface can exhibit either the largest[139] or the smallest[138] thermal resistance within the three-component solid-ligand-liquid interface as depicted in Fig. 3.

Similar to bare solid-liquid interfaces, non-bonded interactions between the ligands and solvents affect the TBC.[139,142] For polar interfaces, electrostatic forces promote the formation of stable hydrogen bonds at the ligand-liquid interface.[143] Stronger hydrogen bonding has been shown to draw organic solvent molecules closer to the interface, resulting in tighter molecular packing. The increased proximity, along with a higher atomic number density of the organic liquid near the interface, facilitates thermal energy transport. The strength of ligand-liquid interactions can be tailored by modifying the chemical composition of the functional groups on the ligand layer.[78,144] Shavalier and Gezelter[145] investigated the influence of ligand-to-solvent hydrogen bonding on heat transfer and demonstrated that PEG-capped AuNPs in water exhibit enhanced thermal conductance. Their analysis of vibrational power spectra revealed an increased population of low-frequency heat-carrying modes (0-70 cm$^{-1}$) for the thiolated PEG. Because of the Bose-Einstein weighting of lower frequency modes, improved thermal transport was observed. Their findings also suggest that solvent penetration and ligand configuration-specifically, the orientational ordering of ligand chains-play crucial roles in interfacial heat dissipation. Alternatively, experimental measurements by Tian et al.[144] demonstrated that the TBC is insensitive to the ligand's chain length, suggesting that interfacial transport at the ligand-water interface is primarily dictated by the chemistry of the terminal groups on Au surfaces exposed to water molecules. These findings, highlighting the influence of chain length and solvent penetration, have been further corroborated by MD simulations, as demonstrated in the work by Stocker and Gezelter, examining thiolate-capped gold surfaces.[146]

Computational investigations on surface ligand coverage have shown that partly covered surfaces enhance the TBC as illustrated in Fig. 7.[136,141] This enhancement is attributed to the increased number of thermal exchange pathways and improved vibrational coupling between the hexylamine ligands-passivated CdSe interfaces and the surrounding hexamine solvent. However, at high surface coverage, a critical turning point is reached at which TBC begins to decrease.[147] This reduction occurs because excessive ligand coverage prevents effective penetration of liquid molecules into the ligand layer. As surface coverage continues to increase, the reduced mobility of liquid molecules within the densely packed ligand layer hinders interfacial heat transfer, leading to diminished TBC. Alternatively, Zhang et al.[100] demonstrate that decorating interfaces with high-coverage polymeric SAMs can significantly enhance the TBC even between materials with considerable vibrational mismatch. Specifically, they reported a 430% increase in the TBC after coating both sides of graphene with 7.14% polyethylene (PE), using polymethyl methacrylate (PMMA) as the surrounding medium. This enhancement was attributed to three key factors: (i) the formation of extended and well-aligned polymer chains within the PE/PMMA blending region, (ii) strong vibrational coupling between PE and PMMA, and (iii) covalent bonding between graphene and PE chains.

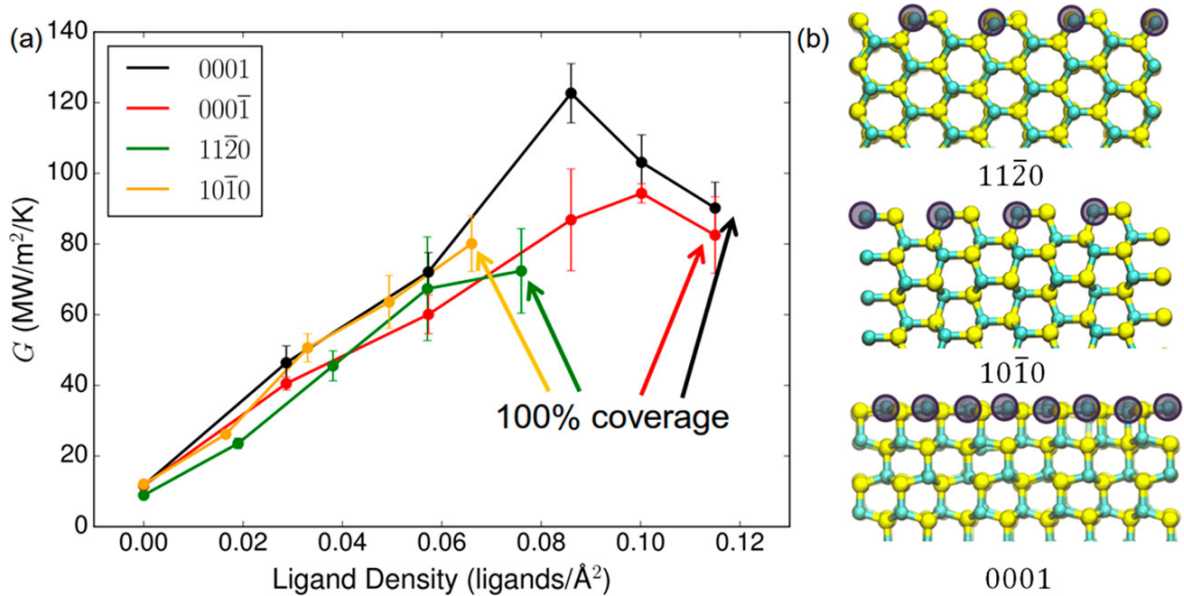

**Fig. 7** Interfacial thermal conductance, *G*, as a function of ligand density on various CdSe surfaces. (this figure has been adapted from ref. 136 with permission from ACS Publications, Copyright 2015).

The mobility of liquid molecules at the solvent-ligand interface plays a complex role in determining the net thermal transport across functionalized solid-liquid interfaces. On one hand, reduced liquid mobility, particularly that of water molecules, has been correlated with enhanced TBC in AuNPs functionalized with various organic ligands.[148] This is because low mobility promotes better alignment of liquid molecules with the ligand layer, facilitating more efficient phonon scattering across the ligand-liquid interface.[135,146] For example, when liquid hexene molecules align with thiolate chains in the ligand layer, the vibrational coupling improves, enhancing thermal exchange. Alternatively, low interfacial liquid mobility also has a downside, as observed when hexene molecules become trapped on thiolated-capped Au surfaces, thereby hindering heat transfer by molecular diffusion.[146] This duality highlights the intricate balance between alignment and mobility in governing interfacial thermal transport.

In addition to mobility, the organization of liquid molecules at the interface and their proximity to both the solid surface and the ligand layer play a critical role in interfacial thermal transport. Liquid molecules in closer proximity to the solid surface and the ligand layer facilitate more effective energy exchange at the interface.[142] However, the role of liquid layering and structuring in thermal transport across functionalized solid-liquid interfaces remains debated. Neidhart and Gezelter[149] found that a higher solvent density peak within the penetration region correlated with an increased TBC. Conversely, Sun et al.[140] observed a weaker dependence of the TBC on liquid layering for gold slabs coated with SAM. They concluded that liquid layering effects are more pronounced in bare solid-liquid interfaces than in functionalized ones. This highlights the nuanced and context-dependent role of liquid structuring in thermal transport across functionalized interfaces.

Analyzing heat transport across functionalized solid-liquid interfaces is inherently complex due to the coexistence of a three-component interface. Unlike bare solid-liquid interfaces, additional factors must be considered when evaluating the solid-ligand-solvent layer. (i) Ligands exhibit some degree of Brownian motion, unlike solid atoms; however, they lack the mobility of liquid molecules and do not diffuse.[137] (ii) The ligand-water interface is not well-defined because water molecules can penetrate the ligand layer.[136,148] This penetration significantly influences the ligands' temperature profile.[148] For bare or hydrophilic ligand-coated interfaces, the temperature profile typically shows a single steep descent at the solid-ligand interface. However, for hydrophobic ligand-coated interfaces, the temperature profile becomes more complex, exhibiting an initial drop at the solid-ligand interface, followed by a plateau along the ligand, and finally, a second drop at the liquid-ligand interface.[148] This intricate behavior underscores the need for detailed analysis to understand thermal transport in such systems.

Computing the TBC at a complex solid-ligand-liquid interface requires a simplified model to account for its intricacies. A common approach involves calculating a global TBC by considering the temperature change from the solid surface to an idealized sharp ligand-liquid interface.[136,138,146] This method reduces the three-component interface into two independent interfaces: the well-defined solid-ligand interface and the diffuse ligand-water interface, which is approximated as a sharp boundary. Alternatively, some authors employ an effective thermal resistance model (the inverse of TBC), which represents the interface as a network of smaller thermal resistances (as illustrated in Fig. 3). These resistances are defined by the discrete temperature jumps observed at different points across the interface.[141,146] This approach allows for a more detailed characterization of thermal transport mechanisms at the interface. Major findings related to TBC modeling efforts and their implications in biomedical fields are summarized in Table 1.

Table 1. Overview of research on TBC across Au-functionalized interfaces pertinent to biomedical applications.

| Source | Methodology | Key Findings | Implications |
|---|---|---|---|
| Shavalier et al[134] | Computational MD simulations | • Surface morphology significantly impacts interfacial heat transport in functionalized AuNPs.<br>• Stronger ligand-water interactions promote better coupling, enhancing TBC at the gold-water interface. | • Ligand selection can improve heat dissipation and drug delivery efficiency.<br>• Tailoring ligand chemistry can fine-tune interfacial heat transfer properties for biomedical applications. |
| Li et al[92] | Computational MD simulations | • Ordered structuring of interfacial water molecules enhances thermal conductivity. | • Advanced interfacial models should incorporate liquid structuring effects for accurate TBC predictions. |
| Tascini et al[56] | Computational MD simulations | • Curved AuNPs exhibit altered vibrational modes that influence interfacial thermal conductance. | • Curvature effects should be considered in AuNP design for targeted heat transport applications. |
| Ge et al[59] | Experimental | • Ligand-to-water hydrogen bonding enhances thermal coupling, enhancing TBC. | • Functionalization strategies should focus on maximizing ligand-water interactions to enhance heat transfer. |
| Hannah et al[136] | Computational MD simulations | • Higher surface ligand density initially improves TBC, but excessive coverage inhibits effective energy transfer. | • Balancing ligand surface coverage is crucial for optimizing thermal response without hindering drug release. |

## 3. Experimental Measurements of Thermal Transport across Solid-Liquid Interfaces

In the past two decades, there has been considerable advancement in the theoretical understanding of interfacial thermal transport across solid-liquid interfaces, which has mainly been fueled by the tremendous progress in the atomistic modeling based on MD simulations of several solid-liquid interfaces,[74,110,113,150–156] and the advances achieved in the analytical description utilizing the phonon theory of liquid thermodynamics.[157–160] Although comparatively there have been much fewer experimental works focusing on understanding thermal transport across solid-liquid interfaces,[70,77,144,161–166] a handful of these contributions have provided crucial validation to the theoretical advancements. For instance, Harikrishna et al.[77] have shown that by varying the terminal of the alkane-thiol monolayers on the gold thin film surface, the thermal conductance values monotonically increased in the range of 60-190 MW m$^{-2}$ K$^{-1}$ as the work of adhesion increased (Fig. 4(a)), and the solid-liquid contact angles spanned from 25° to 118°. The measurements were carried out with the time-domain thermoreflectance technique (TDTR), which utilizes a femtosecond pulsed laser system to monitor (in real-time) the temperature changes on a metallic surface induced by the laser pulse adsorption.[77,167,168] In fact, such pump-probe laser-based techniques are particularly well suited for this application and have been the popular choice for investigating interfacial heat transfer across solid-liquid interfaces.

In the pump-probe-based thermoreflectance techniques, the laser pulses are absorbed by the solid (usually Au thin films deposited on a transparent substrate), and a bidirectional heat flow model is used to back out the thermal boundary conductance across metal-liquid interfaces (as schematically represented in Fig. 8(a)). The popular choice for the liquid has been water and to vary the interfacial adhesion (hydrophobicity), the well-known gold-thiol chemistry (as utilized by Harikrishna et al.[77]) is utilized. However, the use of the traditional thermoreflectance techniques, which have been the current standard for measuring thermal boundary conductance, lacks sufficient sensitivity to accurately quantify the interfacial heat conduction across solid/liquid interfaces.[161]

The experimental insensitivity to solid-liquid TBC, in general, originates from the large thermal resistance posed by the liquids relative to that of the interfacial heat flow.[169] This has been shown quantitatively through calculations of the sensitivity (based on TDTR sensitivity analysis carried out by Costescu et al.[170]) of the typical TDTR signal (representing the ratio between the in-phase and the out-of-phase signals) to the various parameters in the thermal model used to back-out the thermal boundary conductance (Fig. 8(c)). The relative sensitivity of the solid-liquid TBC is significantly lower in comparison to the thermal conductivity of the liquid due to their low thermal diffusivities.

Recently, Tomko et al.[161] highlighted the lack of sensitivity of the typical pump-probe experiments to solid-liquid interfacial heat flow by using gold films in contact with several different liquids. In this work, the pump and probe beams were passed through transparent glass substrates and focused on the surface of gold films in contact with various liquids at the other end. Similar to the conventional approach, a bidirectional heat flow model was used to back out the thermal boundary conductance. The results from the measurements were compared with TBC measured across other gold-substrate interfaces. In this regard, it is instructive to compare the values as a function of the ratio of the longitudinal sound velocities for the two media comprising the interface as shown in Fig. 8(b).

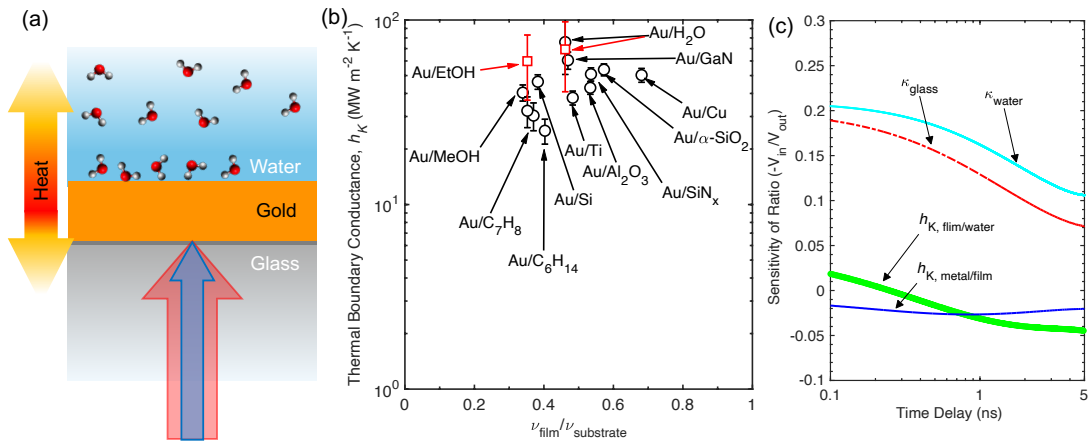

**Fig. 8** (a) Schematic of the three-layer system (usually incorporating glass/Au/water layers) utilized in typical TDTR measurements where the pump and probe beams are incident on the gold surface after passing through the transparent substrate. (b) Measurement of the TBC lower bounds for a thin film $Au/H_2O$ and Au/ethanol interfaces. (c) Sensitivity of the solid-liquid TBC in comparison to the thermal conductivity of the liquid in typical TDTR analyses.

Assuming a simple acoustic mismatch model (AMM) for thermal boundary conductance,[171] increasing overlap of the sound velocities, $\nu$, of the two media is expected to enhance the heat conduction and reduce the temperature drop that occurs at that interface. Although the comparison of the measured values suggested that the TBC across gold-liquid interfaces can be as high as those of other solid-solid interfaces associated with gold films, these measurements represented lower bounds (with the error bars representing a 5% error in the film thicknesses). For these gold-liquid interfaces, it was not possible to obtain a nominal value for the upper bound of the measured TBC with the conventional TDTR technique alone, as the experimental insensitivity to solid-liquid TBC originated from the large thermal resistance posed by the liquids.

Although it has been difficult to accurately determine the TBC with the typical thermoreflectance techniques, the work by Tomko et al.[161] showed that alternative pump-probe experiments to quantify nanoscale energy transport at solid-liquid interfaces can support the traditional measurements and provide the much-needed validities. Namely, they probed the damping of acoustic phonon modes (commonly referred to as picosecond acoustics) in the solid layer upon interaction with the solid-liquid interface. This technique is based on the optical detection of the propagation of acoustic modes through the piezo-optic effect and can provide information on the transmissivities of acoustic phonons across the interface between the solid film and contacting layers.[161] In other words, the ultrafast pump pulse excitation of metal films produces an oscillatory strain wave that travels in the thin film and interacts at the metallic film interface, which attenuates the oscillatory strain, and thus allows for the measurement of the phonon mode transmissivity across that interface.[172–174]

Tomko et al.[161] showed that the transmissivities increased with the increase in the work of adhesion at the solid-liquid interfaces. They further supported their measurements with experiments that monitored the ablation threshold for the various samples, which served as a metric for changes in thermal transport at the gold-liquid interfaces. More specifically, the ablation threshold for gold thin films in contact with different liquids was shown to correlate well with TDTR measurements of TBC across the Au-liquid interfaces.[175] Note, that the ablation threshold is a quantitative measure of the minimum laser fluence required to remove mass from the thin gold films in the experiments. Notably, in Tomko et al.'s work, this threshold was highly dependent on the TBC between the Au films and contacting layers, where the ablation threshold increased linearly with increasing TBC. By correlating this linear increase with the ablation thresholds measured at the various gold-liquid interfaces, it was shown that the TBC across the Au-liquid interfaces could be determined with lower uncertainties as compared to the regular TDTR measurement analysis procedure with large uncertainties.[161] Therefore, although current thermoreflectance techniques lack the required sensitivity to accurately determine TBC across solid-liquid interfaces, a combination of such pump-probe experiments can provide a founding platform for advancing our fundamental understanding of interfacial heat flow in these systems.

## 4. Computational Modeling of Solvated Gold Nanoparticles

Molecular dynamics (MD) is a powerful computational tool used to model atomic-level interactions using a classical physics framework. In the field of heat transfer, MD facilitates the calculation of the thermal conductivity and TBC. Furthermore, the atom-level resolution offered by MD enables a deeper understanding of the underlying mechanisms of thermal transport. Achieving meaningful results from MD simulations requires atomic structures to ideally be derived from experimental data or principles of crystallography to accurately represent a solid. Additionally, it is of paramount importance to select or develop a suitable set of force fields (FFs) that faithfully capture the interactions within the system and align with the specific research objectives. This Section provides guidance for proper MD modeling of functionalized gold-water interfaces, which is critical for advancing the application of plasmonic nanoparticles in the biomedical field. Accordingly, this section is organized into three parts: Section 4.1 reviews various MD methods for calculating interfacial thermal transport at solid-liquid interfaces. Next, Section 4.2 delves into relevant literature covering MD foundations on modeling different components of bare and functionalized gold water interfaces. Finally, Section 4.3 offers a concise overview of the models in two subsections: Subsection 4.3.1 summarizes the extensive work on Au-water interactions, and Subsection 4.3.2 covers models of thiolate adsorption on Au surfaces, including mathematical representations of the Au-sulfur bond.

**4.1 Molecular Dynamics Methods for Thermal Boundary Conductance Calculations**

The TBC is crucial for characterizing nanoscale heat transfer, and MD simulations offer several methods to calculate it. Rajabpour et al.[176] compared four primary MD techniques for evaluating the TBC at nanoparticle-water interfaces: The transient non-equilibrium molecular dynamics (TNEMD) method using both lumped capacitance and finite internal resistance models, the steady non-equilibrium molecular dynamics (SNEMD) method, and the equilibrium molecular dynamics (EMD) approach. TNEMD exists in two versions: (i) the lumped capacitance model, which assumes negligible internal conduction resistance; and (ii) the finite internal resistance model, which accounts for internal temperature gradients. The lumped model closely mirrors transient experimental setups but struggles with accurately defining temperatures due to rapid cooling. Conversely, the finite internal resistance model captures internal temperature gradients but risks inaccuracies when nanoparticles are smaller than phonon mean free paths.

In the lumped capacitance model, the NP is selectively heated at a temperature $T_i$ and then allowed to cool down in a large constant temperature fluid reservoir at $T_\infty$. The NP's temperature decay is fitted using Eq. (3):

$$T_{NP}(t) = T_\infty + (T_i - T_\infty)e^{-\frac{t}{\tau}} \quad (3)$$

where $T_{NP}(t)$ is the NP's temperature at time $t$, and $\tau$ is the thermal relaxation time. The TBC is then calculated as

$$G = \frac{C_{NP}}{A\tau} \quad (4)$$

where $C_{NP}$ is the NP's heat capacity and $A$ is its surface area. In contrast, the finite internal resistance model accounts for spatial-temporal variations of temperature inside a NP, while assuming that the macroscopic heat conduction equation (Eq. (5)) governs this problem:

$$\frac{\partial T_{NP}(r,t)}{\partial t} = \alpha \nabla^2 T_{NP}(r,t) \quad (5)$$

where $T_{NP}(r,t)$ is the local temperature at radius $r$ and time $t$, and $\alpha$ is the thermal diffusivity of the NP. By tracking the temperature of discrete spherical shells over time and fitting to analytical or numerical solutions, the TBC is calculated from the NP's surface boundary condition, which is a convection or Robin boundary condition. Both methods enable quantitative evaluation of the TBC, but the choice between them depends on the particle's Biot number, where the TBC is not known a priory. The differences between the two TNEMD approaches, manifested as less accurate temperature relaxation profiles and subsequently less precise estimates of interfacial thermal conductance,[177] for the interface between an alkane nanodroplet and water.

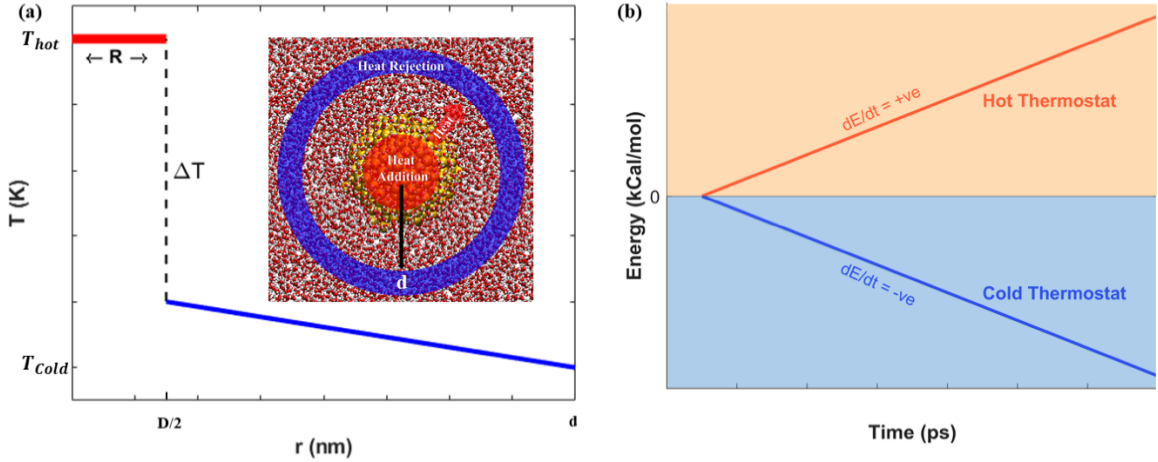

**Fig. 9** SNEMD simulation setup for AuNP-water system: (a) Temperature distribution from the heated AuNP to the cooled water at the boundaries of the simulation domain. (b) Time-dependent accumulation of energy input into the AuNP and extraction of water; the linear slope indicates the established heat flux.

The SNEMD method is widely used to determine the TBC by imposing a continuous temperature gradient across an NP-fluid system. In this approach, the system is divided into distinct thermal regions: a central NP or solid region is maintained at a higher temperature (heat source), while distant fluid regions are kept at a lower temperature (heat sink), see the inset in Fig. 9(a). The temperature gradient can be either created using thermostats[176] or heat input/output regions[85]. Once a steady-state temperature profile is established (Fig. 9(b)), the TBC is calculated using Eq. (6):

$$G = \frac{q}{A \Delta T} \qquad (6)$$

where $q$ is the heat transfer rate across the solid-liquid interface, $A$ is the interfacial area, and $\Delta T$ is the temperature drop at the interface, see Fig 9(a). If the thermostat method is used, $q$ can be computed from the rate of energy added to (or removed from) the thermostatted region as:

$$q = \frac{1}{\Delta t}\frac{\Delta E}{A} \tag{7}$$

where $\Delta E$ is the total energy added or extracted over time $\Delta t$, see Fig 9(b). To obtain $\Delta T$, the solvent's and NP's temperature profiles are spatially averaged, and a linear fit is performed on the solid regions away from the interface; the extrapolated temperature discontinuity at the interface yields $\Delta T$.

A detailed understanding of interfacial thermal coupling can be obtained from EMD methods.[68] Unlike non-equilibrium simulations, EMD does not generate temperature gradients; instead, it relies on the fluctuation-dissipation theorem to relate heat transport properties to equilibrium energy exchanges between the solid and liquid regions. Once the system reaches thermal equilibrium at a given temperature $T_0$, the EMD method uses fluctuations in the interfacial heat power to calculate $G$ via the Green-Kubo formalism[178]:

$$G = \frac{1}{Ak_B T_0^2} \int_0^\infty \langle P(t)P(0)\rangle dt \tag{8}$$

where $A$ is the interfacial cross-sectional area, $k_B$ is the Boltzmann constant, $T_0$ is the equilibrium temperature of the system, $P(t) P(0)$ is the instantaneous heat power/flux across the interface, and $\langle \rangle$ denotes an ensemble average for the time autocorrelation function of the fluctuating heat power. Barrat and Chiaruttini[179] noted that the conventional Green-Kubo relation shown in Eq. (8) is strictly valid only in the thermodynamic limit, where the volumetric heat capacity $c_v \to \infty$. Since MD simulations inherently deal with finite systems, they proposed a modified Green-Kubo expression for solid-liquid interfaces in finite domains, calculating $G$ from the long-time integral of the interfacial heat power autocorrelation function.[179]

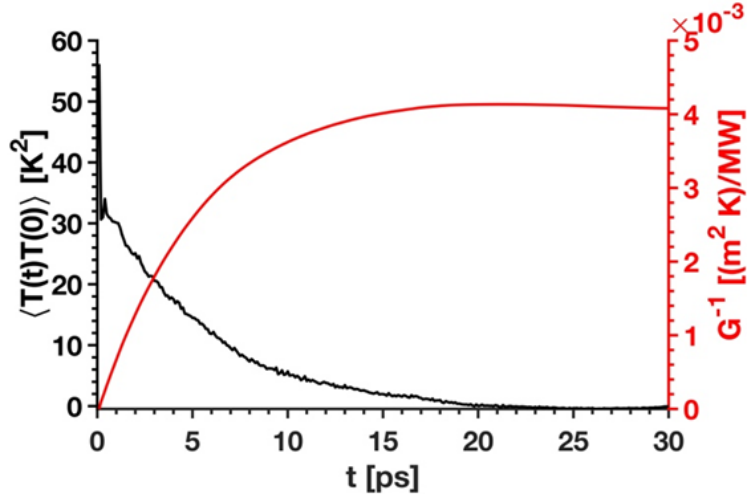

**Fig. 10** Equilibrium MD results for the AuNP and interfacial water. Left Y-axis: Temperature difference autocorrelation function over time. Right Y-axis: Interfacial thermal resistance (Inverse of TBC). (Reprinted with permission from ref. 85. Copyright 2023 AIP publishing).

Recently, Rajabpour and Volz[180] developed a new Green-Kubo expression for the TBC of finite systems, in which $G$ is derived by integrating the time autocorrelation function of the instantaneous temperature difference between the solid and the surrounding liquid:

$$\frac{1}{G} = \frac{A}{k_B T_0^2} \int_0^\infty \langle \Delta T(t) \Delta T(0) \rangle dt \qquad (9)$$

where $\Delta T(t)$ is the instantaneous temperature difference between the solid and the first solvation shell of the surrounding liquid. To enhance statistical reliability, the temperature autocorrelation function is averaged over time and multiple configurations as shown in Fig. 10. While EMD avoids artifacts introduced by artificial temperature gradients, it requires long simulation times and careful statistical averaging to obtain converged results due to the inherently noisy nature of equilibrium fluctuations. Nevertheless, it is particularly suited for systems where applying external gradients would be physically unrealistic or introduce unwanted nonlinearities.

The SNEMD method requires large and unphysical temperature gradients to compute a TBC with minimal statistical noise. Similarly, it may introduce nonlinear temperature artifacts due to thermostating. In contrast, the EMD approach avoids such artifacts by relaying on equilibrium temperature fluctuations, making it more suitable for evaluating intrinsic interfacial properties. Rajabpour et al.[176] systematically compared four MD-based techniques and reported that while all predicted TBCs within the same order of magnitude, SNEMD and EMD differed by less than 10%. In contrast, TNEMD with a lumped thermal resistance approximation underestimated the TBC by approximately 25% compared to EMD, whereas TNEMD with a finite internal resistance overestimated it by a similar margin. This discrepancy is likely due to the transient nature of TNEMD, where the NP is initially heated and allowed to cool, making it challenging to define a precise interfacial temperature and, therefore, accurately quantify the TBC. More recently, the TBC calculations produced by the three different MD techniques (lumped capacitance TNEMD, SNEMD, and EMD) were reported by Paniagua et al.[85] for the TBC across Au-water interfaces. It was reported that although the EMD method provides less artificially imposed conditions on the system, the implementation is computationally expensive and prone to instabilities during the computation of the time autocorrelation function. Therefore, due to its smaller temperature gradients and affordability, the SNEMD method was appropriate to compare the TBC of the different interfaces.[85] Thus, careful consideration of these factors is crucial for accurate TBC evaluation in molecular-scale thermal transport investigations.

**4.2 Computational Modeling of Solvated Functionalized AuNPs**

Gold is a highly valuable material with diverse applications in fields such as materials science, nanotechnology, and catalysis.[181] Consequently, the literature on MD modeling of Au-based systems is extensive. Common FFs for modeling gold include Lennard-Jones (LJ),[182] Morse,

Embedded Atom Method (EAM), and Quantum Sutton-Chen (QSC). Pairwise FFs, such as LJ and Morse, offer computational efficiency and are parameterized to match key physical properties. LJ potentials effectively capture lattice constant, cohesive energy, bulk modulus, and free surface energy of metals.[182–186] In contrast, Morse potentials, featuring tunable parameters, are commonly used to represent covalent bonds[187,188] but can also represent non-covalent interactions, though LJ potentials perform better in this role.[189] Morse FFs accurately reproduce the bulk properties of Au,[183] and Au surfaces, providing useful modeling of thiol adsorption on Au substrates[190] (see Subsection 4.3.2 for more details).

While pairwise FFs are widely used in MD simulations, they have notable limitations, especially in modeling metallic solids. Specifically, pairwise FFs cannot adequately represent electron delocalization and charge transfer, essential for accurately modeling metallic bonding.[185,191] To address these shortcomings, more sophisticated FFs like EAM have been developed.[192,193] For Au, EAM has successfully described the properties of both bulk systems[194] and NPs,[195–197] but at higher computational costs compared to pairwise FFs. Another sophisticated approach is the QSC FF, which also accurately describes the metallic interactions.[198] However, both the EAM and QSC FFs do not account for polarization effects, limiting accuracy in simulating the interfaces between a metal substrate and a polar adsorbate. To address this gap, Bhattarai et al.[89,199] developed the density readjusting embedded atom method (DR-EAM), explicitly incorporating polarization effects by assigning partial charges to atoms, thereby adjusting valence electron densities. Interestingly, this contribution demonstrated that polarization effects had minimal influence on the interfacial vibrational characteristics and thermal transport properties at metal-water interfaces. Thus, the choice of FF for modeling Au should be guided by specific interaction properties, system complexity, and computational efficiency needs.

Water is widely used in solvated AuNP models due to its similarities to biological fluids[200] and well-documented properties. Its models typically use Coulomb's law for electrostatics and LJ potentials for dispersion and repulsion. Charges may be placed at atomic cores or on dummy sites, with the LJ term frequently reserved for oxygen-oxygen interactions. Water models are distinguished by factors such as interaction point count, rigidity or flexibility, and polarization inclusion. Three-site models, such as SPC (Simple Point Charge) and TIP3P (Transferable Intermolecular Potential 3-Point), are among the earliest water models[201–204]. These models apply partial charges and LJ parameters to the oxygen atom and the two hydrogen atoms. While their simplicity allows for efficient large-scale MD computations, they regard the water molecule as rigid and nonpolarizable, limiting their capacity to effectively represent temperature-dependent characteristics. Four-site models, notably TIP4P and its improved versions such as TIP4P/2005, add a dummy atom to refine the electrostatic distribution [201,205]. Although they outperformed three-site models in terms of predicting structural properties and phase diagrams, they lacked explicit polarizability, limiting their accuracy in strong electric fields or inhomogeneous situations. To further enhance the accuracy, five-site models like TIP5P and six-site models have been developed [206,207], but they are less frequently employed due to high computational demand.

Sirk et al.[208] conducted a comprehensive comparison of the thermal conductivities of rigid and flexible water models using MD. Flexible models, such as TIP3P/Fs, SPC/Fw, and SPC/Fd, demonstrated 15-25% greater conductivities than their rigid counterparts due to having more degrees of freedom, allowing for more efficient energy transmission. On the other hand, thermal conductivities for rigid models like SPC, SPC/E, TIP3P-Ew, and TIP4P-Ew ranged between 0.776 to 0.816 W/mK, with an average of 0.799 W/mK, while the experimental value is 0.609 W/mK at 300 K. Recent advancements include polarization-corrected and flexible water models.

Polarization enhancements better represent water properties in polarizable environments.[204] Rigid water models use position constraints to treat bonded interactions implicitly, while flexible models capture anharmonic O–H bond stretching. Researchers demonstrated that the choice of water model has a considerable impact on the strength and heat flux dependency of TBC at the nanoscale Au-water interface [209]. They found that due to increased phonon coupling at low frequencies, the rigid TIP3P model provides higher and more consistent TBC values. The flexible TIP3P model, on the other hand, yields somewhat lower and temperature-dependent TBC, with increased conductance at greater heat fluxes.

Thiolates (R-S⁻) are sulfur-based radicals bonded to organic groups, commonly found in thiols (R-SH) and as ligands in Au-SAM interfaces or thiolate-protected AuNPs. Their interactions in functionalized metallic nanostructures are often modeled using all-atom (AA) and united-atom (UA) FFs. Examples include the OPLS-AA and OPLS-UA FFs[132,148,210,211], the TraPPE-UA FF,[149,212] and FFs developed for organic molecules and proteins, such as CHARMM,[213] AMBER,[214] and GROMOS.[215] All-atom FFs explicitly account for interactions between each atom intramolecularly, while united-atom FFs group certain atoms, such as hydrogens bonded to carbon atoms, into a single interaction center. Both FF types incorporate bonded (covalent) and non-bonded (van der Waals and electrostatic) interactions. Bonded interactions include terms for bond stretching, angle bending, dihedral angle torsion, and improper torsion, providing a detailed framework for modeling thiolates.

Unlike traditional FF, which need individually parameterized models for Au, water, and thiol ligands, and frequently rely on combining rules to address cross-interactions, ReaxFF offers a unified reactive framework capable of expressing any pertinent interactions under a single parameter set.[216] ReaxFF can dynamically describe bond formation and breaking, allowing for reliable modeling of Au-S chemisorption, interfacial water structuring, and ligand reconfiguration under temperature gradients.[216,217] This avoids the need to mix diverse non-reactive force fields (e.g., EAM for Au, SPC for water, and OPLS-AA for ligands), minimizing compatibility issues and enabling more adaptable and chemically consistent simulations under different circumstances. Table 2 summarizes the FFs for solvated functionalized AuNPs modeling.

Table 2. Summary of FFs for solvated functionalized AuNPs models.

| Force field | Feature | Pros / Cons |
|---|---|---|
| **Au modeling** | | |
| Pairwise FFs (LJ, Morse) | Simple two-body potentials | • Computationally efficient<br>• Poor at capturing metallic bonding |
| Many-body FFs (EAM, QSC, DR-EAM) | Include many-body and/or polarization effects | • Better accuracy for metals<br>• Higher computational cost |
| **Water modeling** | | |
| Rigid models (SPC, TIP3P, TIP4P/2005) | Fixed bond lengths and angles | • Efficient for large simulations<br>• Limited accuracy for temperature-dependent properties |
| Flexible models (TIP3P-Fs, SPC/Fw) | Allow bond vibrations | • Captures thermal effects better<br>• Costlier than rigid models |
| **Ligands modeling** | | |
| All-atom FFs (OPLS-AA, CHARMM, AMBER) | Explicitly model each atom; detailed bonded and non-bonded interactions | • High accuracy for organics<br>• Tedious parameterization and high cost |
| United-atom FFs (OPLS-UA, TraPPE-UA) | Groups non-polar H atoms with heavy atoms for efficiency | • Lower computational demand<br>• Less detail for H bonding |
| Unified Option: ReaxFF | Reactive FF handling bond formation/dissociation across Au, water, and ligands | • Chemically consistent unified modeling<br>• Relatively computationally expensive |

## 4.3 Computational Modeling of Interfaces

Modeling interfaces in MD is inherently more complex than modeling bulk materials or isolated molecules. Interfaces involve the coexistence of two dissimilar materials, requiring detailed analysis of molecular adsorption onto the interface and diffusion across the interface. Interfacial FFs are typically optimized to capture surface properties such as wetting behavior, surface tension, and adsorption energies.[186,218] However, FF parameters tailored for interface interactions are not always readily available. In such cases, simple approaches, such as empirical combination rules for LJ FF parameters, are often employed. Nonetheless, for interfaces exhibiting both physisorption and chemisorption, non-bonded interactions alone lack the precision required to accurately describe interfacial structures. For strongly interacting interfaces, such as Au-S systems, the FF must be developed with a focus on the underlying adsorption physics to ensure accuracy.

The literature on ligand-solvent interfaces is relatively limited. Consequently, it is common practice to model ligand-solvent non-bonded interactions using Lorentz-Berthelot combining rules, which merge parameters from all-atom FFs (see Section 4.2) with the non-bonded interactions of water models.[149,181,212,219] For instance, in systems of functionalized metallic nanoparticles immersed in water, several works have employed the SPC/E water model combined with the OPLS FF.[148,212,220] The body of work on MD models for Au-water interfaces is extensive, and concise details are provided in the following Subsection.

### 4.3.1 Metal-Solvent Interfaces

Most MD models of metallic-fluid interfaces define the interfacial interactions between the metal and the adjacent fluid using a truncated 12-6 LJ FF, with Lorentz-Berthelot combining rules commonly applied to determine FF parameters. Berg et al.[221] examined the limitations of using simple LJ FFs and combination rules for describing interfacial interactions in Au-water systems. Their findings revealed that LJ FFs poorly replicate adsorption energy curves obtained through density functional theory (DFT) simulations, particularly when compared to more sophisticated pairwise FFs, such as Buckingham or Morse. Furthermore, Berg et al.[221] were unable to identify a suitable set of FF parameters that could successfully use combination rules to match DFT-calculated adsorption energies, highlighting the need for more accurate approaches to model such interfaces.

Alternatively, FF parameters for metal-fluid interfaces have been determined by optimizing the interfacial properties of interest, such as experimental contact angles,[129] adsorption energies,[218] surface tension,[186] or DFT-derived adsorption energy curves.[221,222] However, FF parameters optimized using different methods often exhibit discrepancies when calculating interfacial properties.[108,118] Additionally, these parameters are rarely transferable across different systems. It is then imperative to shift the modeling strategy from generic empirical FFs to new FFs based on the physics and chemistry of specific interfaces. ReaxFF,[216] a bond order-based force field capable of accounting for reaction energy barriers, chemical bonding, and non-bonded interactions of solid-water interfaces, is needed to model Au-water interactions for a better understanding of thermal transport. There is substantial research focusing on MD models for thiolate adsorption on Au surfaces and mathematical representations of the Au-sulfur bond. These models are summarized in the following Subsection.

### 4.3.2 Gold-SAM Interfaces

Despite extensive research on SAM-Au interfaces and their numerous applications, the precise arrangement and configuration of adsorbed thiolates have been a topic of debate for decades.[223–229] Notably, most investigations on SAM-Au interfaces focus on systems with an exposed Au (111) surface.[230] Early electron diffraction characterizations suggested that thiols organize into a hexagonal ($\sqrt{3} \times \sqrt{3}$) R30° lattice,[231,232] commensurate with the underlying Au (111) surface, as depicted in Fig. 11(a). Theoretical and computational investigations of this ($\sqrt{3} \times \sqrt{3}$) R30° model propose that thiolates preferentially occupy threefold-coordinated hollow sites, twofold-coordinated bridging sites, or positions directly above Au surface atoms,[224–226,233–238] as illustrated in Fig. 11(b).

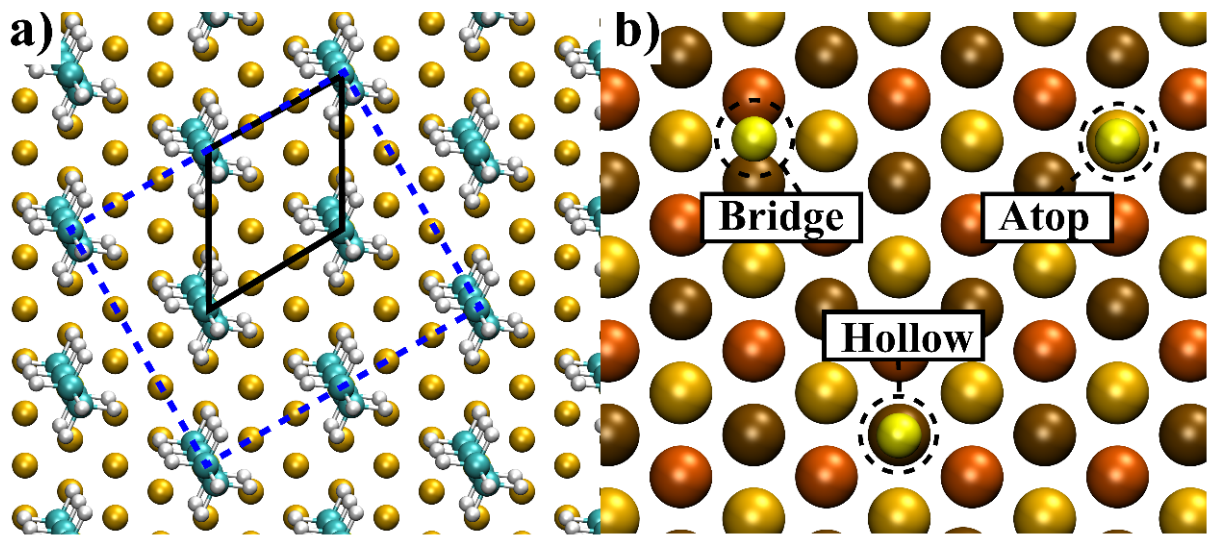

**Fig. 11** Adsorption of sulfur on the Au (111) surface. a) Early experiments identified that adsorbed sulfur atoms organized in a hexagonal (√3 × √3) R30° lattice above the Au (111) surface (solid black line), and the existence of a c (4 × 2) superlattice (dashed blue line). Gold, teal, and white spheres represent Au, carbon, and hydrogen atoms, respectively. b) Theoretical and computational calculations concluded three possible adsorption sites for the sulfur atoms: hollow sites, bridge sites, and atop sites. Gold, orange, and ochre spheres represent Au atoms in the outermost, 2$^{nd}$ and 3$^{rd}$ layer of an Au (111) surface, respectively, while yellow spheres represent adsorbed sulfur atoms.

The traditional (√3 × √3) R30° lattice arrangement has been increasingly challenged over the years. X-ray measurements revealed the existence of a centered (4 × 2) superlattice derived from the (√3 × √3) R30° structure. This centered (4 × 2) superlattice consists of four atoms, where two adsorbed thiols are equivalent, and the other two occupy distinct lateral and vertical positions relative to the underlying Au (111) surface[239] (see Fig. 11(a)). Furthermore, recent scanning tunneling microscopy (STM) visualizations have identified the presence of Au adatoms at the SAM-Au interface.[223,240–242] These observations demonstrated that thiolates form RS-Au-SR complexes on a reconstructed Au (111) surface, particularly at low thiol coverage.[223,241,243,244]

STM has revealed that the Au-SAM interface is better described as a complex assembly of thiolates bonded to Au adatoms, rather than thiolates directly adsorbed onto an atomically flat Au (111) surface.[240] Additionally, the Au-SAM interface has been shown to exhibit dynamic behavior, including the diffusion of thiolates on the Au surface and the exchange of adsorption sites.[245,246] For instance, DFT combined with *ab initio* MD simulations have demonstrated that adatom-based structures can emerge from the reconstruction of the interface when a system initially organized under the ($\sqrt{3} \times \sqrt{3}$) R30° lattice model is allowed to relax.[243] However, STM measurements confirming the presence of RS-Au-SR complexes have not yet been achieved for intermediate to full thiolate coverage on Au-SAM interfaces.[247,248]

The presence of adatoms is particularly prominent in AuNPs, which exhibit a core-shell-like structure, where the gold core atoms are surrounded by shell-like Au adatoms bonded to the sulfur head groups of thiolates.[249] The formation of RS-Au-SR complexes was first observed in AuNPs by Jadzinesky et al.[250], who identified dimeric and monomeric RS-Au-SR staples (see Fig. 12), with their distribution and ratios varying depending on the nanoparticle size. More recent research has suggested the existence of trimeric SR(-Au-SR) staples,[251] bridging thiolates,[252,253] and cyclic -Au-SR structures.[251] However, distinguishing the latter from the more abundant dimeric and monomeric RS-Au-SR staples remains challenging.[254]

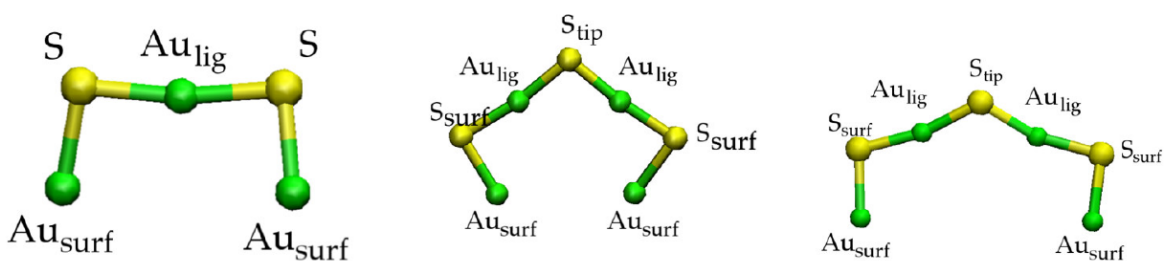

**Fig. 12** Monomeric and dimeric RS-Au-SR staples in the shell of AuNPs[255]. The green and yellow spheres represent Au and sulfur atoms, respectively. (this figure has been adapted from ref. 255 with permission from ACS Publications, Copyright 2016).

DFT has been widely employed to investigate the structural properties of thiolated Au surfaces. However, the omission of dispersion forces in some DFT simulations has introduced uncertainties in the adsorption behavior of thiols on Au surfaces,[256] with results showing a strong dependence on the chosen functional.[225] Several investigations have focused on elucidating the fundamental aspects of AuNP-ligand bonding. For instance, Reimers et al.[257] used DFT calculations to show that the stability of sulfur-stabilized AuNPs arises from local gold–sulfur covalent interactions rather than from complete electron shell closure. Complementing this, Tang and Jiang[258] systematically evaluated the binding strengths of various ligands on gold surfaces, revealing clear trends where bulky N-heterocyclic carbenes and alkynyl groups form particularly robust bonds. In a similar vein, Pensa et al.[248] reviewed the complexity of the sulfur–gold interface, highlighting the multiple coordination modes and dynamic surface reconstructions that challenge simplified models of AuNP functionalization.

Methodological advances have also played a crucial role in deepening our understanding of these systems. Fusaro et al.[259] combined DFT with solvent models to accurately predict adsorption energetics and ligand exchange processes, while Berg et al.[221] optimized force fields for water–gold interactions, improving the reliability of MD simulations in reproducing experimental observations. Several reviews[260–262] have synthesized these computational approaches, addressing challenges such as weak intermolecular forces, multiscale phenomena, and the dynamic nature of the bio interface. Nevertheless, the high computational cost of DFT makes it impractical to model the length and time scales found in heat transfer processes in complex Au-SAM interfaces. Consequently, MD simulations have become the preferred approach for investigating thiol-protected AuNPs[263–267] and flat Au-SAM interfaces.[268,269] Unfortunately, FF parameters available for describing the Au-S bond are often developed with a focus on specific adsorption models. As a result, these FFs are frequently non-transferable between Au-SAM and thiolated AuNP systems, or even between AuNPs of different sizes.[270] However, reactive FFs, such as ReaxFF, may be trained using the high-fidelity insights from DFT calculations, which include surface reconstructions, bonding configurations, and adsorption energies to make them more accurate and transferable across different Au-S interfaces.

An early attempt to model the Au-sulfur interaction in SAMs utilized an LJ FF.[271,272] However, this approach had notable limitations, including the oversimplification of the Au surface as flat, as the Au-S potential energy was treated as dependent solely on the perpendicular distance from the surface. Furthermore, LJ FFs have been shown to inadequately capture the chemisorption behavior of sulfur on Au surfaces.[189] To address these limitations, Perstin and Grunze[273] developed a modified LJ FF that incorporated a surface corrugation function and explicitly accounted for Au-S-C angle bending in thiolates, providing a more accurate representation of the Au-SAM interface.

Morse FF parameters have been developed as an alternative to better describe the chemisorption of sulfur on Au surfaces.[274–276] However, these parameters are often tailored to specific sulfur binding sites, such as the threefold hollow site on Au (111) surfaces,[274,275] limiting their transferability to other Au surface configurations or sulfur binding models. To address this limitation, more sophisticated functional forms and models have been proposed to accommodate different binding site scenarios. For instance, Longo et al.[277] developed a modified Gupta FF to account for Au vacancies and adatoms at the Au-sulfur interface, while the GolP FF[278] was designed to accurately represent the atop binding site for sulfur. These advancements offer improved flexibility in modeling complex Au-sulfur interfaces.

The non-transferability of Au-SAM FFs poses a significant challenge in modeling thiolate-protected AuNPs. Due to the highly anisotropic surface of AuNPs, FFs developed for specific binding sites on flat Au surfaces cannot be directly applied. To address this, robust elastic network models with optimized force constants have been developed to describe the AuNP core-shell structure (including adatoms) and Au-sulfur interactions.[148,210] In particular, ReaxFF[279–282] is well-suited to capture complicated chemisorption and surface reconstruction events in thiolated AuNPs due to its capability to dynamically represent bond dissociation and formation. Similarly, Pohjolainen et al.[270] developed a transferable all-atom FF for thiolated AuNPs, with parameters optimized to account for various staple units, enabling more versatile modeling of complex Au-sulfur interfaces. Table 3 summarizes the key insights from DFT and experimental studies on Au-S interfaces.

**Table 3.** Summary of key insights from DFT and experimental studies on thiolated Au interfaces.

| Aspect | Key Findings | Method/Reference |
|---|---|---|
| SAM Lattice Structure | Hexagonal ($\sqrt{3}\times\sqrt{3}$)R30° and c(4×2) superlattices observed depending on coverage and reconstruction [231,232] | LEED, X-ray diffraction, STM |
| Sulfur Adsorption Sites | Occupation of hollow, bridge, and atop sites [224–226,233–238]; sensitive to the local environment and surface relaxation | DFT, STM |
| Presence of Adatoms | RS-Au-SR staples (monomeric, dimeric) were observed especially at low coverage and in AuNPs [255] | STM, XPS, DFT |
| Ligand Binding Stability | Covalent Au-S bonding dominates; dispersion effects important; binding strength varies by ligand type [257] | DFT studies with functional-dependent results |
| Surface Reconstruction | Thiol adsorption induces surface rearrangements and dynamic site switching [248] | *Ab initio* MD, STM |
| Limitations of Traditional FFs | Fixed-site or LJ-based FFs fail to reproduce adsorption flexibility and reconstruction | MD modeling comparisons |
| Advanced Modeling Strategies | Use of Morse [274–276], Gupta [277], GolP [278], ReaxFF [279–282] FFs, and DFT-calibrated approaches for specific binding modes | Computational FF development studies |

## 5. Summary and Future Outlook

AuNP-enabled therapy has the potential to advance the medical field, particularly in therapeutic applications, such as drug/gene delivery and photothermal therapy. This review highlights their unique properties, including plasmonic behavior, biocompatibility, and versatile surface functionalization, which collectively and combined with precise thermal modulation, could deliver highly-localized drug release and tissue ablation therapies. The consideration of rDA reactions in drug delivery systems further exemplifies how the thermal properties of AuNPs can be leveraged for spatiotemporal therapeutic interventions. A key focus has been the interfacial thermal transport mechanisms of AuNPs, particularly at gold-water interfaces, where functionalized ligands significantly influence heat dissipation. Computational works have provided critical insights into the interplay between ligand chemistry, interfacial water structuring, and nanoscale thermal transport, shedding light on the modulation of the TBC.

Despite these advancements, challenges persist in optimizing solvated functionalized AuNP systems with adequate spatiotemporal temperature control in complex biological environments. Such optimization requires a proper description of interfacial heat transfer across functionalized Au-water systems, which is governed by a complex interplay of multiple mechanisms. In the experimental field, TDTR measurements have dominated in obtaining measurements of TBC across solid-liquid interfaces, though they have critical limitations in terms of sensitivity. Improving or developing new experimental techniques with higher sensitivity to solid-liquid TBC is essential to enhance the accuracy of measurements. For example, combining TDTR with alternative methods like picosecond acoustics and ablation threshold measurements can provide a more comprehensive understanding of TBC and reduce uncertainties. Given the current experimental limitations, it is not surprising that computational modeling remains the favored approach for understanding heat transfer across functionalized Au-water interfaces.

In the computational realm, MD models have strongly dominated over *ab initio* efforts due to the high computational cost of the latter, which limits their applications to the large size and time scales required to model complex functionalized Au interfaces. Numerous MD works have extensively investigated the individual mechanisms governing interfacial thermal transport in functionalized Au-solvent interfaces, including the atomic interaction strength, the role of hydrogen bonding, vibrational mismatch, and the mobility and molecular organization of the liquid phase near the interface. However, the isolated descriptions fail to fully capture the coupled nature of interfacial thermal transport. Additionally, notable discrepancies exist in the literature on the role of mobility and interfacial structuring in defining the TBC of functionalized interfaces. These discrepancies arise from the use of different frameworks to characterize the liquid layering effect, overlooking the broader molecular organization of the interfacial liquid. Notably, for bare Au-solvent interfaces, the role of interfacial structuring is better understood, as TBC calculations have been explained using the liquid depletion layer parameter $\delta$.

To further explore interfacial thermal transport, a continuum model incorporating AuNPs and their surrounding medium can be developed. This model must integrate granularity, temperature-dependent chemistry, and interfacial liquid property variations into a temperature-dependent TBC calculation. By capturing the transient thermal response of AuNPs and the surrounding water, the model would ensure interfacial temperature continuity, providing a comprehensive framework for nanoscale heat transfer analysis. Moreover, a quantitative analysis of energy exchange contributions at each sub-interface could provide a more precise characterization of interfacial heat transfer in functionalized Au-solvent interfaces. Developing comprehensive transient thermos-chemical models will lay the foundation for investigating complex biomedical applications, such as drug delivery systems using rDA reactions.

Nevertheless, as demonstrated in MD simulations, accurately describing interfacial atomic interactions is crucial for predicting interfacial thermal transport. Precise parameterization of FF parameters is essential for reliable reproducing interfacial properties. Reactive force fields, such as ReaxFF, offer a realistic representation of metal-liquid interfaces by capturing bond formation and dissociation; however, their implementation involves high computational costs that must be carefully managed. In this context, *ab initio* models provide a cornerstone for deciphering the complex surface chemistry of functionalized Au interfaces. Additionally, the advent of machine learning FFs,[283] trained to provide the accuracy of *ab initio* methods in an efficient classical framework, has opened new avenues for modeling of interfaces.[284]

Addressing these challenges will unlock the full potential of AuNPs, establishing them as indispensable tools for advancing biomedical technologies. These efforts will enhance the precision and effectiveness of therapeutic interventions while contributing to the broader vision of personalized medicine, where treatments can be tailored for maximum efficacy and minimal side effects. As a final remark, it is important to note that several points discussed throughout this review are not exclusive to functionalized Au-solvent systems and can be extended to other solid-liquid interfaces of interest with appropriate considerations. Nevertheless, for the sake of brevity and coherence, the discussion primarily addressed functionalized Au-water interfaces, as they are the main system of interest in biomedical applications of plasmonic NPs and the central focus of this review.

**Author Contributions**

**Md Adnan Mahathir Munshi:** Data curation (experimental and numerical part, equal), Formal analysis (experimental and numerical part, equal), Investigation (experimental and numerical part, equal), Writing-original draft (experimental and numerical part, equal)

**Emdadul Haque Chowdhury:** Data curation (experimental and numerical part, equal), Formal analysis (numerical part, equal), Investigation (numerical part, equal), Writing-original draft (numerical part, equal), Writing -review & editing (equal)

**Luis E. Paniagua-Guerra:** Data curation (experimental and numerical part, equal), Formal analysis (experimental and numerical part, equal), Investigation (experimental and numerical part, equal), Writing-original draft (experimental and numerical part, equal), Writing -review & editing (equal)

**Jaymes Dionne:** Data curation (experimental part, equal), Formal analysis (experimental part, equal), Investigation (experimental part, equal), Writing-original draft (experimental part, equal)

**Ashutosh Giri:** Data curation (experimental part, equal), Formal analysis (experimental part, equal), Investigation (experimental part, equal), Writing-original draft (experimental part, equal), Writing -review & editing (equal), Funding acquisition (equal), Supervision (equal)

**Bladimir Ramos-Alvarado:** Data curation (experimental and numerical part, equal), Formal analysis (experimental and numerical part, equal), Investigation (experimental and numerical part, equal), Writing-original draft (experimental and numerical part, equal), Writing -review & editing (equal), Funding acquisition (equal), Supervision (equal)

## Data availability

No primary research results, software, or code have been included and no new data were generated or analyzed as part of this review.

## Conflicts of interest

There are no conflicts to declare.

## Acknowledgments

This work was supported by the National Science Foundation, USA (Award number: 2430793).